\def\({\left(}
\def\[{\left[}
\def\l\{{\left\{}
\def\){\right)}
\def\]{\right]}
\def\r\}{\right\}}
\def\what{\widehat}
\def\raw{\rightarrow}
\def\bA{\bar A}
\def\am{\bar A_{\rm max}}
\def\cf{{\cal F}}

\def\etal{{\sl et al.} }
\def\ratio#1#2{{{#1}\over{#2}}}
\def\KH{Kelvin-Helmholtz }
\def\3d{three-dimensional }
\def\2d{two-dimensional }
\def\be{\begin{equation}}
\def\ee{\end{equation}}
\def\bef{\begin{figure}}
\def\ef{\end{figure}}

\def\DXDYCZ#1#2#3{\left({\partial#1\over\partial#2}\right)_{#3}}
\def\lapp{\mathbin{\raise2pt \hbox{$<$} \hskip-9pt \lower4pt
\hbox{$\sim$}}}
\def\gapp{\mathbin{\raise2pt \hbox{$>$} \hskip-9pt \lower4pt
\hbox{$\sim$}}}
\documentclass{aa}
\usepackage{graphicx}
\begin{document}

   \thesaurus{02.01.1; 02.08.1; 02.09.1; 02.19.1; 11.10.1}
   \title{Kelvin-Helmholtz instability in three dimensional radiative
jets}

   \author{M. Micono\inst{1,2} \and  G. Bodo\inst{2} \and S. Massaglia\inst{1} \and 
           P. Rossi\inst{2} \and  A. Ferrari\inst{1,2} \and  R. Rosner\inst{3}
          }

\institute{Dipartimento di Fisica Generale dell'Universit\`a,
Via Pietro Giuria 1, I-10125 Torino, Italy (surname@ph.unito.it) \and
Osservatorio Astronomico di Torino, Strada Osservatorio 20, 
10025 Pino Torinese, Italy (surname@to.astro.it) \and
Department of Astronomy \& Astrophysics, University of Chicago, Chicago IL 60637, 
USA (surname@oddjob.uchicago.edu)}

   \offprints{M. Micono}

   \date{Received; accepted }

\maketitle

\begin{abstract}

The analysis of the stability properties of astrophysical jets against
Kelvin-Helmholtz (or shear-layer) instabilities plays a basic role
in the understanding the origin and physical characteristics of these
objects. Numerical simulations by Bodo et al.\ (1998) have shown that the 
three-dimensional non-linear evolution of KH instabilities in supersonic
jets is substantially faster than in the two-dimensional case, leading to a
cascade of modes towards smaller scales and a very effective mixing and
momentum transfer to the ambient medium. On the other hand, Rossi et al.\
(1997) and Micono et al.\ (1998) found, in two dimensions, that radiative
losses tend to reduce and delay mixing effects and momentum transfer to
the ambient medium. In this paper, as a logical next step, we investigate the
effects of radiative losses on the stability of 3D supersonic jets, 
assuming that the internal jet density is initially lower, equal and higher
than the ambient medium, respectively. We find that light and equal-density
radiative jets evolve in a qualitatively similar fashion with respect to the
corresponding adiabatic ones. Conversely, we note substantial differences 
in the evolution of heavy jets: they remain more collimated and do not
spread out, while the momentum gained by the ambient medium stays within
$\sim 5$ jet radii.

\keywords{Hydrodynamics --
                instabilities -- shock waves -- Galaxies: jets
               }
   \end{abstract}
\section{Introduction}

Jets that originate from young T~Tauri stars in giant molecular clouds and
from Active Galactic Nuclei propagate, in a collimated fashion, up to
distances $\sim 100-1000$ times their average radii. The question of their
endurance against \KH instabilities is a crucial one for understanding the
phenomenology of these objects. This issue has been addressed by many
authors first in the linear limit (see Birkinshaw 1991 for a review) and,
in  recent years, through the use of 2D and 3D numerical simulations 
(see Ferrari 1998 for a review). In particular, recently, Bodo et al.\
(1998) have examined the differences between \2d and \3d jet evolution,
showing that \3d jets undergo faster evolution, enhanced mixing and larger
jet momentum depletion. The temporal duration of the shock-dominated phase 
of the \KH instability evolution is strongly reduced, shocks that form are
weaker, and the direct association between structures observed in
astrophysical jets and shocks originating as a consequence of the nonlinear
development of the \KH instability is more problematic. Moreover, the rapid
disruption that \3d jets  undergo raises the question as to how jets can
survive for the long scale lengths shown by observations. Although
relativistic bulk velocities may be a stabilizing factor in extragalactic
jets, the stability problem remains crucial for the nonrelativistic jets
associated with Young Stellar Objects (YSOs), and radiation might be an
important ingredient in this respect.

The interest in radiation effects on the development of \KH instabilities in
supersonic jets was motivated by the observation of line-emitting knots in
YSO jets. Linear analyses of the stability of radiative jets by Massaglia
et al. (1992) and Hardee \& Stone (1997), showed that radiation losses
typically reduce the growth rates of perturbations. Extending the analysis
to nonlinear amplitudes, Rossi et al.\ (1997) and Micono et al.\ (1998)
found that radiative cooling has noticeable effects on the spatial and
temporal evolution of \2d jets; in particular, the action of cooling leads
to a longer duration of the initial linear stage and of the shock-dominated
stage of the instability; shocks that form appear weaker and the post-shock
temperatures smaller; the importance of the mixing phase is reduced and in
some case, depending on the jet density, it is totally absent. Studying this
problem, Downes \& Ray (1998) found a somewhat contrary result with respect
to Rossi et al.\ (1997): from their temporal simulations they concluded that
cooling enhances the momentum transfer to the ambient medium, increasing the
level of mixing. They suggested that the differences in the two studies might
be due to the different geometries adopted: Cartesian in their case, and
cylindrical in Rossi et al.\ (1997).

The goal of the present work is to study the effect of radiative losses on
the evolution of the \KH instability in a three-dimensional jet, with
particular focus on the issues of mixing and momentum transfer in a
system which is not subject to any superimposed symmetry. Moreover, with
explicit reference to jets from YSOs, we aim to verify whether the action of
cooling is strong enough to prevent a fast disruption of \3d jets, which
must survive for the length and time scales that are observed. Specifically,
we investigate the behavior of dense, light and equi\-dense jets, and analyze
the effects of increasing the amount of energy lost through radiation by
changing the value of a radiative control parameter which is directly related
to the ratio between the cooling and dynamical time scales.

The plan of the paper is the following: in Section 2 we examine the physical
problem and the equations used; the numerical scheme is described in Section
3; Section 4 summarizes the results of our previous studies on the
stability of adiabatic jets; the results of our simulations are discussed in
Section 5. Finally, a summary and the conclusions are given in Section 6.

\section {The physical problem}

We study the evolution of a 3D fluid jet, subject to radiative losses;
the equations regulating our system are thus the ideal fluid equations 
for mass, momentum and energy conservation:

\be
{{\partial \rho} \over {\partial t}} + \nabla \cdot (\rho {\bf{v}}) = 0\,, \\
\label{fluid1}
\ee
\be
\rho {{\partial {\bf v }} \over {\partial t}} + \rho ({\bf v} \cdot
\nabla){\bf v } = -\nabla p\,, \\
\label{fluid2}
\ee
\be
{{\partial p} \over {\partial t}} + ({\bf {v}} \cdot \nabla)p -
\gamma\ {p \over \rho}\ \left[
{{\partial \rho} \over {\partial t}} + ({\bf {v}} \cdot \nabla) \rho
\right] = 
(\gamma - 1) \ \left( {\it \cal Q} - {\it \cal L} \right)
\label{fluid3}
\ee

\noindent
where $p$, $\rho$, and ${\bf v}$ are pressure, density and velocity,
$\gamma$ is the ratio of the specific heats, and ${\it \cal L}$ and ${\it
\cal Q}$ are respectively the energy loss term and the heating term (assumed
here constant, and equal to ${\it \cal L}$ taken at $t = 0$); the right hand
side of Eq. \ref{fluid3} vanishes in a purely adiabatic case.

We also follow the evolution of the number fraction of the neutral 
hydrogen atoms $f_{\rm n}$:

\be
{{\partial f_{\rm n}} \over {\partial t}} + ({\bf v} \cdot \nabla)  f_{\rm n} = 
n_e [-(c_{\rm i} + c_{\rm r}) f_{\rm n} + c_{\rm r}]  \,,
\label{fne}
\ee

\noindent
where $n_{\rm e}$ is the electron density, and $c_{\rm i}$ and $c_{\rm r}$ 
are the ionization and recombination coefficients, which depend on the
temperature $T$ and on the electron density
$n_{\rm e}$ in the following way:

\be
c_{\rm i} = 1.08 \cdot 10^{-8} \cdot 13.6^{-2} \cdot \sqrt{T}\ 
{\rm e}^{(-157890./T)} \ \ \ \ \ {\rm cm}^3 \ {\rm s}^{-1}
\ee

\be 
c_{\rm r} = 2.6 \cdot 10^{-11} / \sqrt T \ \ \ \ \ {\rm cm}^3 \ {\rm s}^{-1} \ 
\ee
(Rossi et al. 1997).

Finally, we trace the jet material by means of a scalar field $\cal T$
whose evolution is followed by integrating the scalar advection 
equation:

\be
{{\partial \cal T} \over {\partial t}} + ({\bf v} \cdot \nabla)  {\cal T} = 0 \ .
\label{advect}
\ee
Cooling is treated here as in our two previous papers on 2D cooling jets, 
and the reader is referred to Rossi et al.\ (1997) for a detailed
description. The assumed losses are a reasonable approximation for
temperatures $T \lapp$ 35,000 K and shock velocities up to about 80 km
s$^{-1}$; both conditions are generally fulfilled in YSO jets. We recall, in
particular, that we assume here that our jet is composed  of atomic gas 
with solar abundances, and we neglect the formation of molecules and
molecular emission. This assumption is consistent with our choice for the
initial jet temperature, which is well above the dissociation temperature
for the hydrogen molecule. 

\section{The numerical setup}

\subsection{The integration method}

The hydrodynamical equations are integrated numerically using a
three-dimensional version of the Piecewise Parabolic Method code (Colella \&
Woodward 1984). The code was parallelized through the MPI package.
Multidimensionality, as well as the inclusion of radiative losses, 
are achieved through operator splitting: the PPM code solves the Eqs.
\ref{fluid1} - \ref{fne}) and \ref{advect}) in a conservative form;
where non-conservative terms are present, the physical quantities are updated
at the end of the main time step. In addition, the integration time step is
constrained by the Courant condition, and is further corrected so that the
temperature does not vary by more than 10\% in a single step.

\subsection{Integration domain, initial and boundary conditions}

Our 3D domain $\{ (0,D) \times (-R,R) \times(-R,R) \}$, with $D=10\pi a$
and $R=6.7 a$ where $a$ is the jet radius, is covered by a $256\times 256\times
256$ grid, and is described by a Cartesian  coordinate system $(x,y,z)$.

The initial flow structure is a cylindrical jet, with its symmetry axis 
lying along the $x$-direction and defined by $(y=0, \ z=0)$. The initial 
jet velocity is along the $x$-direction. The interface between the jet and
the surrounding material is not a vortex-sheet-like interface, but a
smoothly varying velocity shear layer. The form of the initial jet velocity
profile is thus:

\be
V_x(y,z)=V_0\ {\rm sech}\left[\left({\sqrt{y^2+z^2}} \over a\right)^m \right] \,,
\ee
where $V_0$ is the velocity on the jet axis,  
and $m$ is a parameter controlling the interface layer width; in these
calculations we set $m=8$ which implies a shear layer thickness of
approximately 0.4 jet radii.

The initial structure of the density in our domain is defined by:

\be
{{\rho_0(y,z)} \over \rho_{0,jet}} = \nu - (\nu -1) \ 
{\rm sech}\left[\left({\sqrt{y^2+z^2}} \over a\right)^m \right] \,,
\ee
where $\nu$ is the ratio of the density at $r=\sqrt{y^2+z^2}=\infty$,
to that on the jet axis $(y=z=0)$ at the initial time.

We assume that the jet is initially in ionization equilibrium, and in
pressure equilibrium with its surroundings (with an initially uniform imposed
pressure distribution).

Finally, we impose the following initial conditions for the passive tracer
$\cal T$:

\be
\cal T = \left\{ \begin{array}
{cc} \; 1, & \quad r < 1 \\ \; 0, & \quad r > 1 \,.
\end{array} \right.
\label{initrac}
\ee
so that we are able to follow the jet material during its subsequent
evolution.

This initial configuration is then perturbed at $t=0$; we perturb the
transverse velocities $V_y$ and $V_z$, so that a wide range of modes can
be excited. In order to mainly excite the helical mode, we choose the
following functional form for the transverse velocities:

\begin{eqnarray}
\lefteqn{V_y(x,y,z) = V_{y,0} \ 
{\rm sech}\left[\left({\sqrt{y^2+z^2}} 
\over a\right)^m \right] \times {} } \nonumber\\
&& {} \qquad \qquad \, \times {1\over n} \ \sum_{j=1}^{n} \sin(jk_0x + \phi_j) \,,
\end{eqnarray}

\begin{eqnarray}
\lefteqn{V_z(x,y,z) = V_{z,0} \ 
{\rm sech}\left[\left({\sqrt{y^2+z^2}} 
\over a\right)^m \right] \times {} } \nonumber\\
&& {} \qquad \qquad \, \times {1\over n} \ \sum_{j=1}^{n} \cos(jk_0x + \phi_j) \,,
\end{eqnarray}

\noindent
where $V_{y,0}=V_{z,0}=0.05V_0$ is the amplitude of the initial perturbation,
$\{\phi_j\}$ are the phase shifts of the various Fourier components,
$n = 8$ and $k_0 = 2\pi / D$.

As regards the behavior at the boundaries, we adopt free outflow boundary
conditions at the upper and lower boundaries in the $y$ and $z$ directions, 
by setting the gradient of every variable to zero. The central region of the
domain, $-3a < y < 3a$, $-3a < z < 3a$, is covered uniformly by 150 grid
points in the radial direction (with the jet itself initially occupying 50
grid points at $t=0$); at larger distances from the jet axis, both the $y$
and the $z$ mesh sizes increase accordingly to the scaling laws $\Delta
y_{j+1} = 1.02\Delta y_j$ and $\Delta z_{k+1} = 1.02\Delta z_k$ respectively.
The grid along the $x$ direction is instead uniform, and the boundary
conditions are periodic. This choice implies a temporal approach to the
study of the instability evolution, as opposed to a spatial approach where
inflow and outflow  conditions are applied at the left and right boundary
respectively, along the jet flow direction. We refer to Rossi et al.\ (1997)
and Bodo et al.\ (1998) for a more detailed discussion of boundary
conditions, grid settings and comparison between the temporal and spatial
approaches.  

\subsection{Scaling and parameters}

Our calculations are carried out in non-dimensional form. Distances are
thus measured in units of the jet radius $a$, velocities are scaled
to the {\it isothermal } sound speed $c_{\rm s}$, evaluated on the jet axis and
at the initial time $t=0$. Accordingly, time is then measured in units
of the sound crossing time $t_{\rm cr}$.

The dynamical control parameters for our problem are the Mach number
M($=V_0/(c_{\rm s} \sqrt{\gamma})$) and the density ratio $\nu =
\rho_\infty / \rho_{\rm jet}$. The treatment of cooling introduces two more
parameters in our calculations, namely the initial temperature on the jet
axis $T_0$ and the ratio $\tau_{\rm rad}$ between radiative cooling time
$t_{\rm rad} =  p/[(\gamma - 1 ) {\cal L}]$ and the sound crossing time.
Fixing $\tau_{\rm rad}$ is analogous to fixing the product of the jet
particle density and the jet radius $n_0 a$, which was selected in our
calculations since it has more direct observational  meaning (Rossi et al.\
1997). 

In our simulations, we fixed the Mach number $M=10$ and we selected three
values for  the ratio $\nu$ between the external and the jet material, i.e.
$\nu=0.1$  (overdense jet), $\nu=1$ (isodense jet) and $\nu=10$ (underdense
jet). The initial temperature on the jet axis was always $T_0 = 10,000$ K,
and for the product of the jet particle density and the jet radius we 
selected the value $n_0 a = 5 \times 10^{17}\ {\rm cm}^{-2}$, which is
consistent, for example, with a jet radius $a=5 \times 10^{15}\ {\rm cm}$
and a jet particle density $n_0 = 100 {\rm cm}^{-3}$; however, we recall
that our results are valid for all pairs of values of $n_0$ and $a$
for which their product is constant.

For the equidense jet ($\nu =1$) case we performed a simulation with
the value $n_0 a = 5 \times 10^{18}\ {\rm cm}^{-2}$ in order to study 
the effect of varying the radiative cooling time on the instability
evolution.

In Table \ref{tabpar} we summarize the simulations discussed in the present
paper, with the values of the relevant parameters.
\begin{table}[hbt]

\caption{}

\begin{center}
\begin{tabular}{|c|c|c|c|c|}  \hline

Simulation&$M$&$\nu$&$T_0$& $n_0a $   \\

\hline

A      & 10  & 0.1 & $10^4$ K & $5. \times 10^{17}\ {\rm cm}^{-2}$\\
B      & 10  & 10. & $10^4$ K & $5. \times 10^{17}\ {\rm cm}^{-2}$\\
C      & 10  & 1.  & $10^4$ K & $5. \times 10^{17}\ {\rm cm}^{-2}$\\
D      & 10  & 1.  & $10^4$ K & $5. \times 10^{18}\ {\rm cm}^{-2}$\\

\hline
\end{tabular}
\end{center}
\label{tabpar}
 \end{table}

\section{Results of previous studies}

\subsection{Stages of the instability evolution}

\label{khstages}

Previous studies of the temporal (Bodo et al.\ 1994, 1995, Rossi et al.\
1997) and spatial (Micono et al.\ 1998) evolution of the \KH instability
in  two-dimensional jets allowed us to distinguish four stages of evolution;
they were identified in adiabatic and cooling jets in the spatial and
temporal approach, in slab and cylindrical geometry; their features, as for
example the onset time, the duration, the morphological and physical 
evolution of the jet during each stage, were dependent on the parameters
adopted, although the general trend was common to all studied cases.
We recall in the following the main features of each evolutionary stage:

\begin{itemize}
\item{} {\it  Stage 1}: the unstable modes excited by the perturbation grow
in accord with the linear theory. In the latter portion of this stage, the
growth of the modes leads to the formation of internal shocks.

\item{} {\it Stage 2 }: the growth of internal shocks is accompanied by a
global deformation of the jet, which thus drives shocks in the external
medium. These shocks can carry momentum and energy away from the jet, and
transfer them to the external medium. 

\item{} {\it  Stage 3 }: as a consequence of the shock evolution, mixing
between jet and external material begins to occur. The longitudinal
momentum, which is initially concentrated inside the jet radius, is  
spread over a much larger region by the spreading of the jet material.  

\item{} {\it Stage 4}: in this final state, the jet attains a statistically
quasi-stationary state. 
\end{itemize}

The effects of radiative losses on the instability have been traced by Rossi
et al.\ (1997) and Micono et al.\ (1998). The main differences of this
analysis with respect to the adiabatic case are: i) a longer duration
of stages 1 and 2 of the evolution; ii) a decrease of the shock strength and
of the post-shock temperatures, and iii) a general reduction of the
importance of the mixing phase (depending on the values of $\nu$ and $n_0a$),
with a complete absence of the mixing phase for high values of $n_0a$.

\subsection{2D vs 3D behaviour}

In this section we summarize the main results obtained by Bodo et al.\ (1998)
in their comparative analysis of 2D and 3D adiabatic jets. Bodo et al.\
(1998) adopted a Cartesian geometry and a temporal approach (i.e., periodic
boundary conditions in the direction of the jet axis), and superimposed
helical perturbations onto the initial transverse  velocities (see also
below).

The expected differences are mainly due to two factors: first, there are many
more unstable modes in three dimensions than in two dimensions,  and the
growth rates of the 3D modes can predominate. Second, it is well-known that
turbulence in two and three dimensions differs substantially, especially as
far as the mixing properties are concerned. Both effects -- the presence of
unstable high-wavenumber modes, and the cascade to high wave numbers via
nonlinear interactions -- can lead  to the formation of small-scale
structures that are not observed  in two dimensions.

The main features observed by Bodo et al.\ (1998) in the evolution of
over-dense and under-dense three-dimen\-sio\-nal adiabatic jets, compared to analogous \2d
slab jets, can be summarized as follows:

\begin{itemize}

\item[{1.}] Faster evolution.

\item[{2.}] Absence of a clear separation of stages 2 and 3 of the
instability (as defined in section \ref{khstages}).

\item[{3.}] More rapid development of small-scale structures due to the
growth of linearly unstable higher order (azimuthal number $\le 2$)
fluting modes and, at advanced times, to a cascade to small scales through 
nonlinear turbulent processes.

\item[{4.}] Enhanced material mixing between the jet and the ambient 
medium; more effective
and faster momentum transfer from jet to the ambient medium.

\item[{5.}] Enhanced jet broadening.

\item[{6.}] Similar evolution of the largest-scale structures.

\item[{7.}] Similar asymptotic states (although reached at very different
times).

\end{itemize}

\section{Results}

\subsection{Heavy jets ($\nu=0.1$)}
\label{sectnu01}

\begin{figure*}

{\includegraphics[width=\hsize,bb=20 190 600 700]{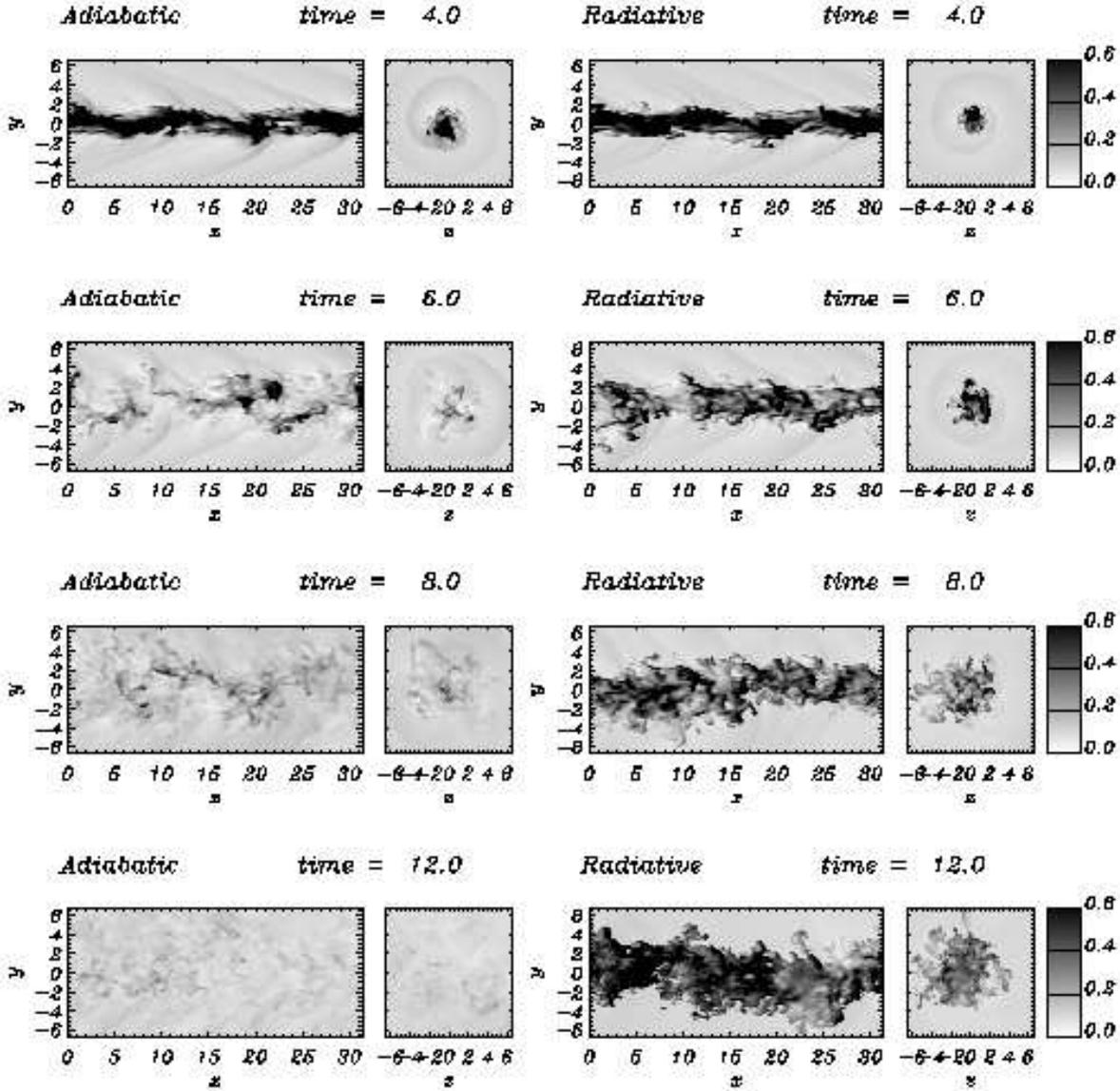}}

\caption{Grey-scale images of the density distribution for heavy ($\nu=0.1$)
adiabatic (left column) and radiative (right column) jets; 
2D cuts at a fixed $z$ ($z=0$) and at a fixed $x$ ($x=5\pi$).}
\label{fig:image_01}

\end{figure*}

\begin{figure*}

{\includegraphics[width=\hsize,bb=20 400 600 690]{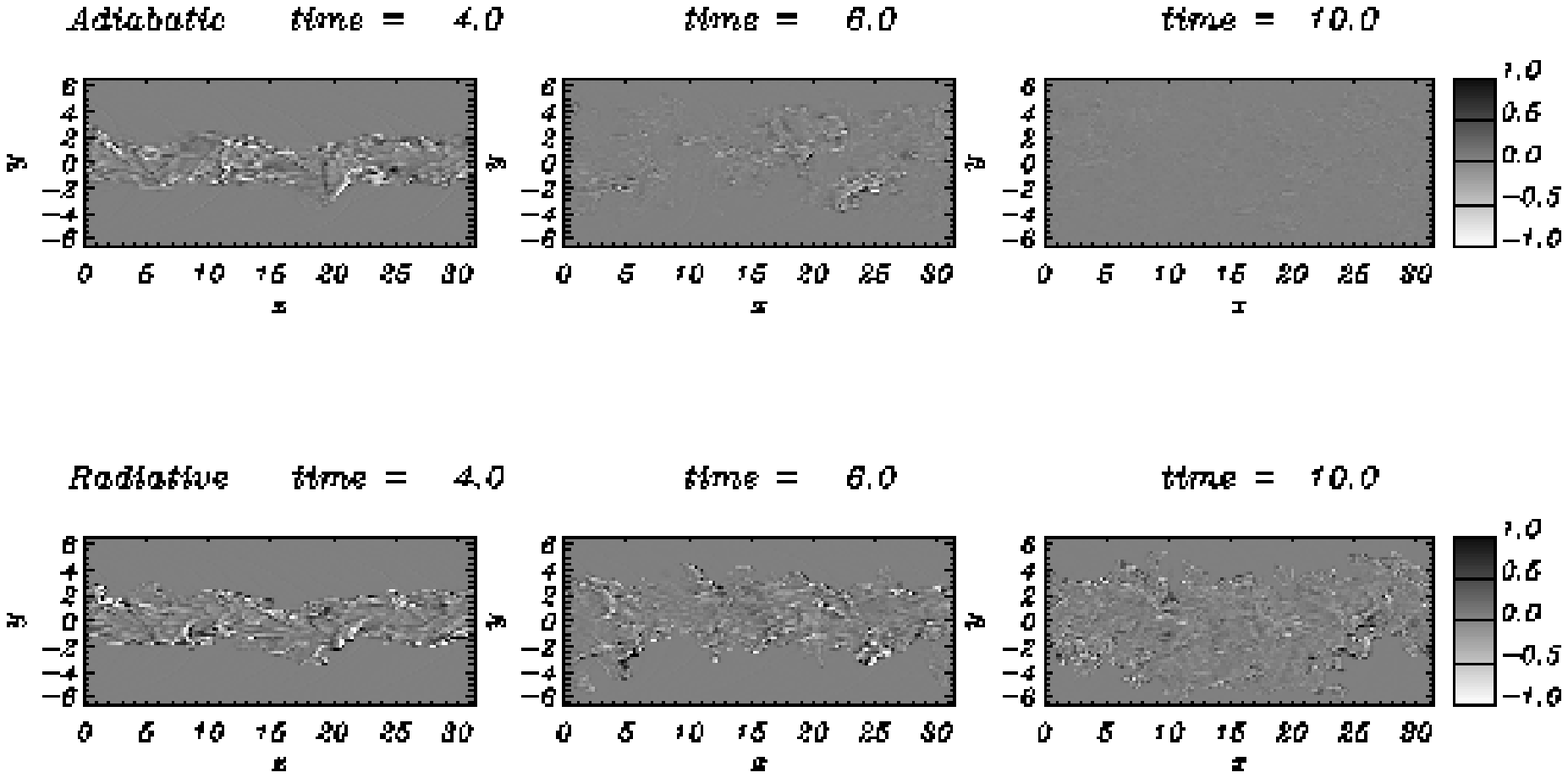}}

\caption{Grey-scale images of two dimensional cuts of the density
Laplacian, in the plane $z=0$
}
\label{fig:lapl}

\end{figure*}

In this section we present the results of the calculations for an adiabatic
jet and a cooling jet with Mach number $M=10$ and density ratio $\nu=0.1$.
We carried out the simulation for $t=12$ units of sound crossing time, since
at this time the adiabatic jet reaches the boundaries in the $y$ and $z$
directions.

In Fig. \ref{fig:image_01} we show grey-scale images of \2d cuts of the jet
density, in the $z=0$ plane (i.e., a longitudinal cut in the $(xy)$ plane in
the middle of the domain) and in the $x=5\pi$ plane (i.e. a transverse
cut in the $(yz)$ plane in the middle of the domain), for the radiative and
adiabatic jets at different stages of the evolution.

At time $t=4$ the situations is similar for the adiabatic and cooling jets: 
the instability has already entered the acoustic phase, with the development
of shock waves driven by the jet into the external medium, in correspondence
to the major ``kinks" caused by the growth of the fundamental helical
mode excited by the imposed perturbation. Acoustic waves are driven
in the external medium, as it can be clearly seen in the upper panels of
Fig. \ref{fig:image_01}. Already at this early stage, a little amount of
external material is being entrained by the jet, as a result of the growth
of higher order fluting modes that produce short scale ripples on the
surface of the jet.

At time $t=6$ the radiative jet is still shock-dominated, although the
disturbances induced on the jet surface have grown in  amplitude and the
amount of entrained ambient medium has increased. In contrast, the
adiabatic jet appears to have fully entered the mixing phase; no trace of
the coherent large-scale deformations induced by the fundamental helical \KH
mode can be discerned. As the evolution proceeds, the adiabatic jet appears
to be completely mixed with the ambient medium ($t=8$, Fig.\
\ref{fig:image_01}), while the jet subject to radiative losses maintains its
collimation up to the latest time in our calculations, $t=12$; at this time
both jets have reached stage 4 of the instability evolution, i.e., a
statistically ``quasi-stationary" stage. The comparison between the
adiabatic and radiative jets in Fig. \ref{fig:image_01} also conveys the
impression that the adiabatic jet has become much wider than its radiative
counterpart. To analyze this effect in more detail, we have computed the
radial distribution of velocity averaged over the longitudinal and
azimuthal directions and we have defined the jet radius as the radius at which
this averaged velocity drops to one half of its maximum value (found at $r =
0$). Fig.\ \ref{fig:emr01}c) shows the evolution of this radius as a function
of time; we see from the figure that in fact the adiabatic jet widens
much more than the radiative one.   

In Fig. \ref{fig:lapl} we show a two-dimensional cut of the Laplacian of the
density in the plane $z=0$ (the Laplacian of the density is a valuable tool
to mark the presence of shock waves in the flow, and is widely used in
gas dynamics).\ \ Upon examining this figure, we can follow the same
evolutionary path described above. The Laplacians at time $t=2$  are not
shown since no significant feature can be detected: both jets are in fact
still in the linear phase (stage 1) of the evolution. When the perturbations
steepen into shock waves, the jet enters the acoustic phase ($t=4$). The
duration of the acoustic phase is longer for the radiative jet, where strong
shock waves survive up to $t=10$; for the adiabatic jet, instead, shock waves
are damped by the vigorous mixing that occurs already during the acoustic
stage: after time $t=6$, no shocks can be  detected in the flow, and
turbulent mixing dominates.

\begin{figure}[htbp]
{\includegraphics[width=\hsize,height=9.6cm,bb=90 90 570 720]{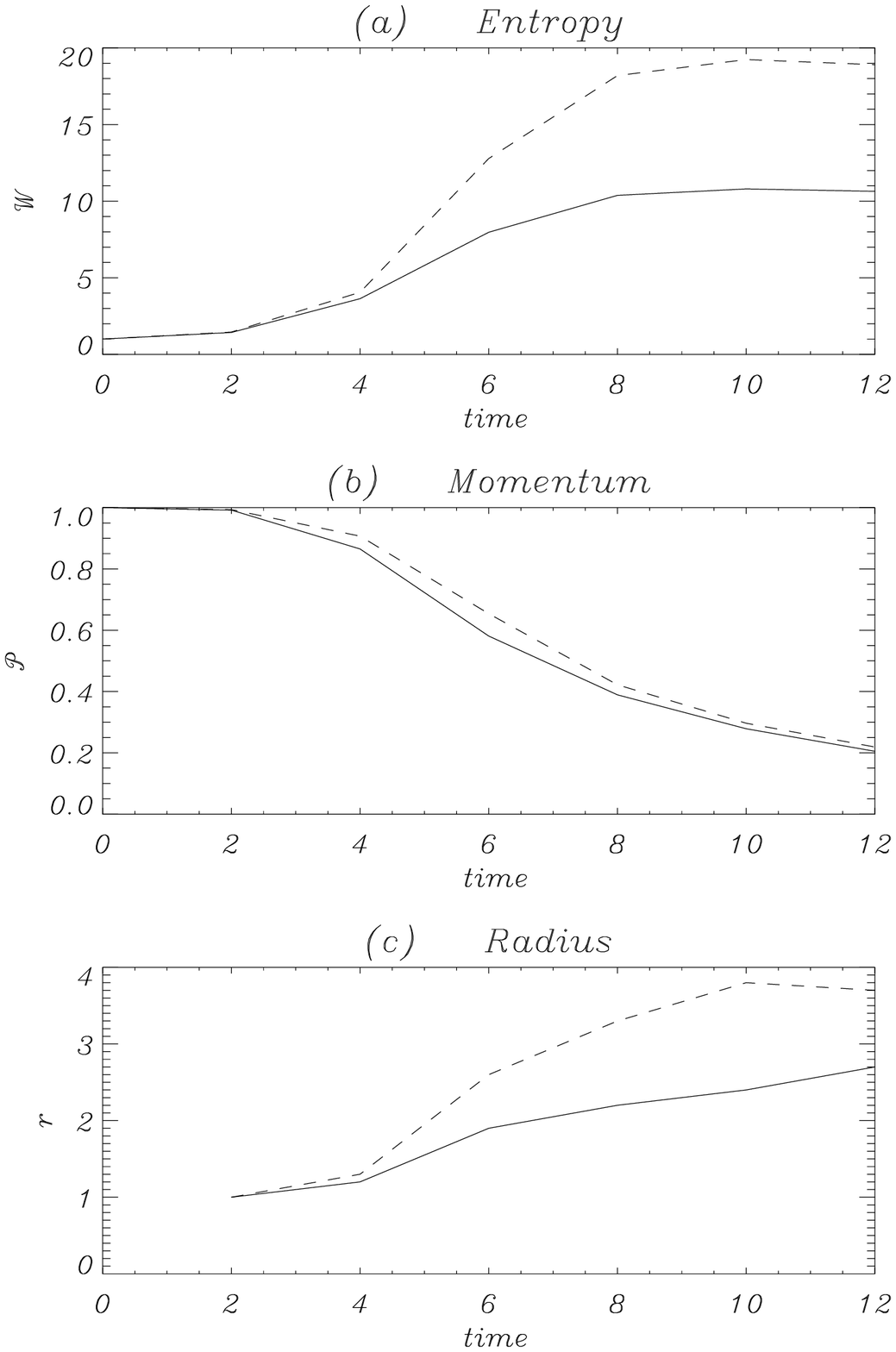}}
\caption{Panel (a) shows a plot of the tracer entropy for the radiative 
(solid line) and adiabatic (dashed line) jets as a function of time (for the
analytical and physical definition of tracer entropy see Bodo et al.\ 1995); 
panel (b) shows a plot of the jet momentum for the radiative (solid line)
and adiabatic (dashed line) jets as a function of time. Panel (c) provides
a plot of the temporal evolution of the jet radius for the radiative (solid
line) and adiabatic (dashed line) cases . All the plots are for the heavy
jet case ($\nu=0.1$).}
\label{fig:emr01}
\ef

\begin{figure*}
{\includegraphics[width=\hsize,bb=80 400 570 770]{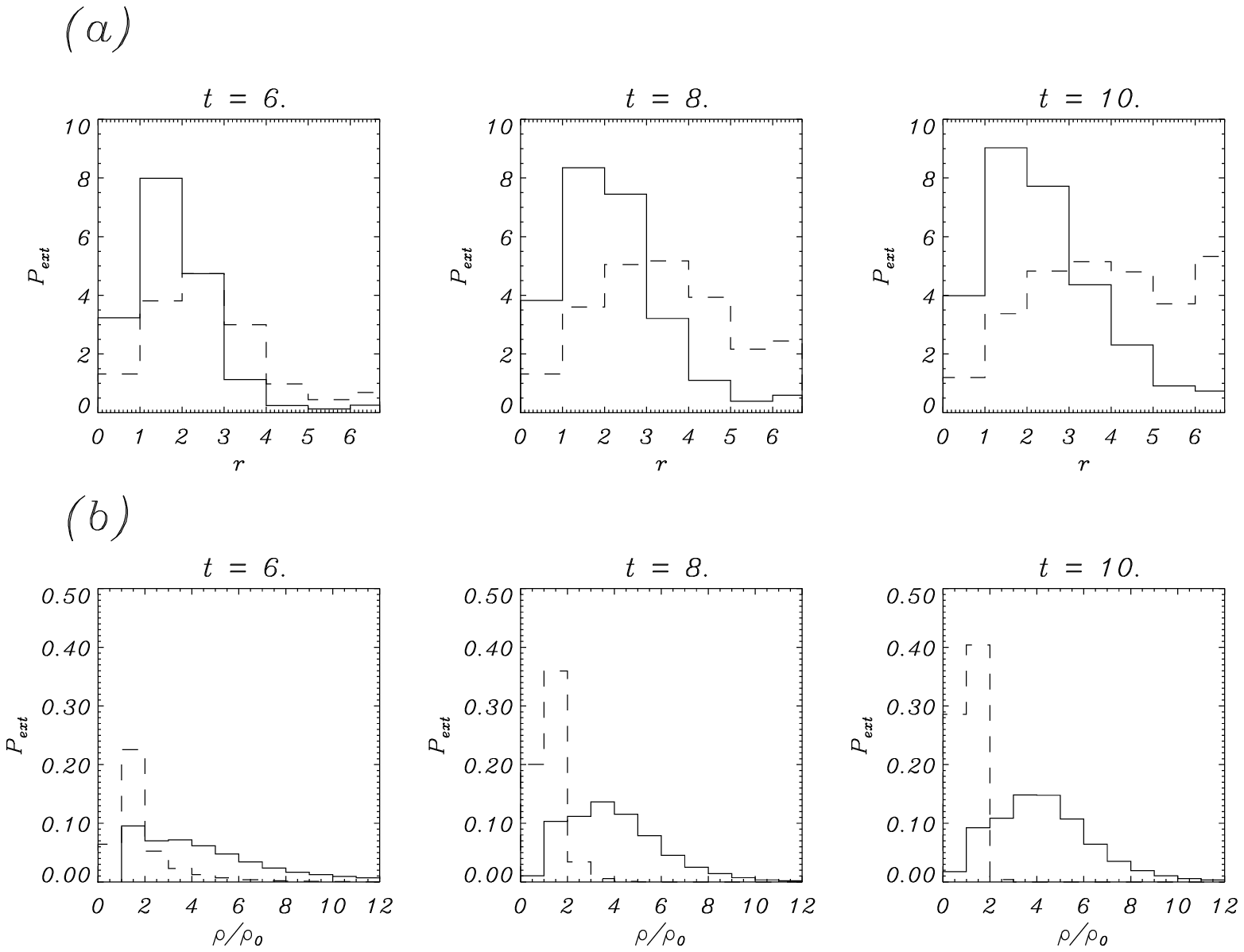}}
\caption{Panel (a): ambient medium momentum as a function of the 
distance from the jet axis. Panel (b): ambient medium momentum versus the
corresponding value of the density. In both panels we plot the trends
referring to heavy radiative  (solid line), and adiabatic (dashed line) jets
at different times. }
\label{fig:mom_distr_01}
\end{figure*}

In panel (a) of Fig.\ \ref{fig:emr01} we plot the tracer entropy $\cal W$
vs.\ time, for the adiabatic (dashed line) and radiative (solid line) dense
jets. $\cal W$  measures the departure of the tracer's distribution from the
initial form (${\cal T} = 1$ for the jet material and ${\cal T} = 0$ for
the external material; see Bodo et al.\ (1995) for the mathematical 
definition of $\cal W$). Up to time $t=2$ the jet is still in the linear
phase of the evolution, and the entropy is almost constant, since the tracer
distribution does not differ substantially from the initial one. As time
elapses, the jet goes through stages 2 and 3 of the evolution, during which
the entropy grows, reaching a quasi-constant value at the end of the
evolution.
 
In Fig. \ref{fig:emr01}b) we plot the momentum of the radiative (solid line)
and adiabatic (dashed line) jets, normalized to the initial values, as a
function of time. In both cases, the jets start losing momentum early in
time; the final momentum loss is dramatic, since at time $t=12$
radiative and adiabatic jets have lost almost 80\% of their initial momentum.
Fig.\ \ref{fig:emr01}b) shows that the amount of momentum lost by the
radiative and the adiabatic jets is similar; the momentum loss is slightly 
greater for the radiative jet. This result seems to be in contradiction with
the behavior described previously: if we assume that the quantity of
momentum lost is a valid indicator of the instability effects, then we would
expect that the radiative jet is equally or more unstable than the
adiabatic jet; on the other hand, we saw that radiative cooling acts
to reduce the development of mixing and to conserve the collimation of the
jet for longer times. In fact, as we have seen above, the jet widening in
the adiabatic case is larger than in the radiative case; one could then 
expect that the amount of entrained material would be much larger in the
adiabatic case, and correspondingly the residual jet momentum much lower.
This apparent discrepancy can be explained if we analyze how momentum is
lost by the jet in the two cases. 

The rate of momentum transfer is almost similar for the cooling and 
adiabatic jets,
but the radiative jet accelerates only the ambient medium located very near 
the jet itself; in contrast, the adiabatic jet expands and entrains
material over a much larger radial extent, accelerating it until at the end
of our calculations all the ambient medium in our computational domain has a
non-zero velocity in the longitudinal direction; this trend appears clearly
in Fig.\ \ref{fig:mom_distr_01}a), where the amount of momentum acquired by
the external medium at different times is shown as a function of the distance
from the initial jet axis.

\begin{figure*}
{\includegraphics[width=\hsize,bb=20 270 600 730]{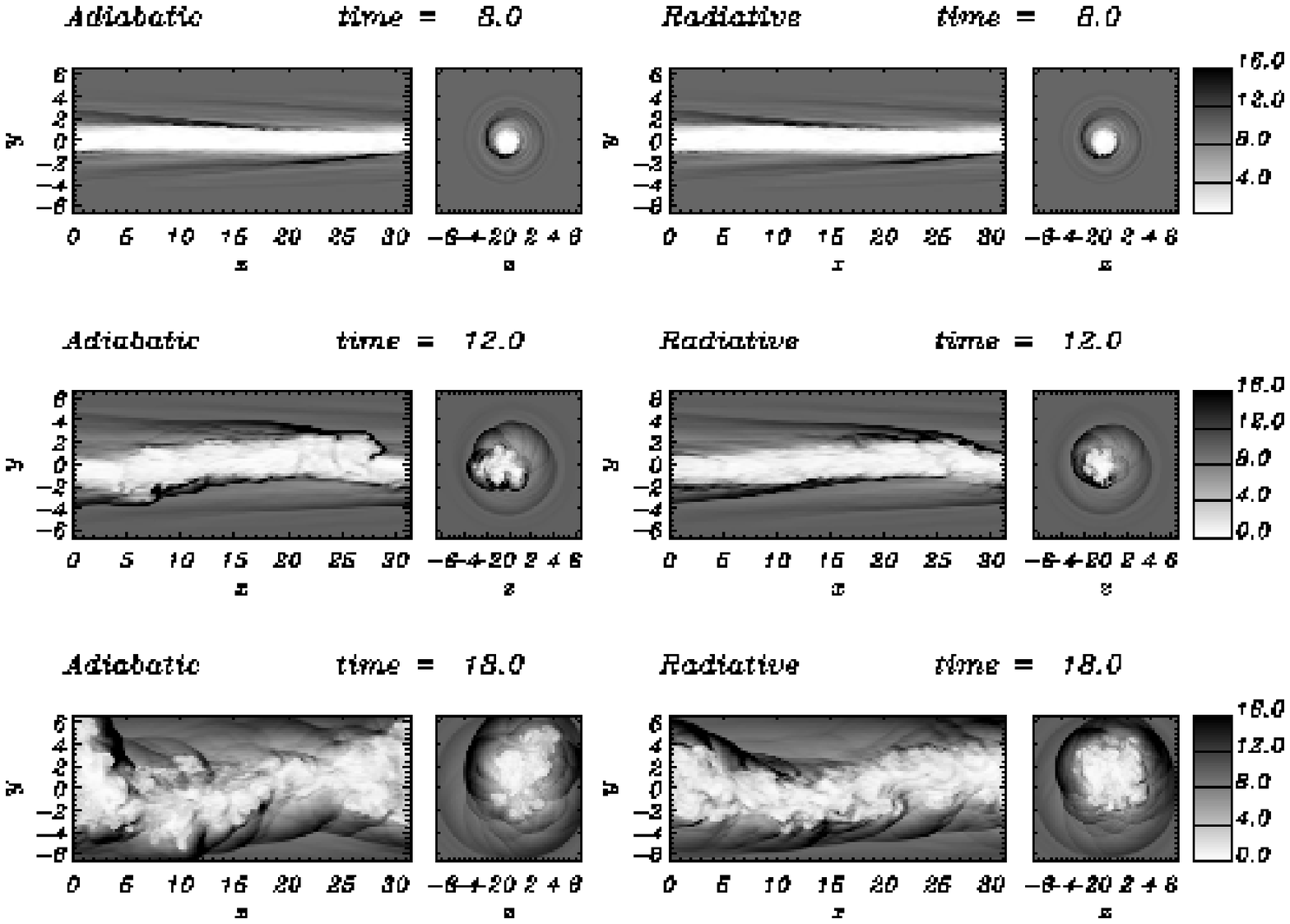}}
\caption{Grey-scale images of the density distribution for light ($\nu=10$)
adiabatic (left column) and radiative (right column) jets; 
2D cuts at a fixed $z$ ($z=0$) and at a fixed $x$ ($x=5\pi$).}
\label{fig:image_10}
\end{figure*}

The typical velocities attained by the accelerated ambient medium in the two
cases are similar (for example, the maximum velocity of the material with
${\cal T} < 0.1$ at t=10 is $\sim 6.8$ for both cases), but the adiabatic jet
accelerates a larger radial {\sl volume} of ambient material. A similar
amount of transferred momentum is thus achieved in the two cases only if in
the radiative case {\sl denser} ambient material is accelerated. That this
is the case is demonstrated by Fig.\ \ref{fig:mom_distr_01}b), which
shows the distribution of momentum in the external medium versus the
corresponding value of the density for the radiative and adiabatic jets: we
characterized each volume of ambient medium by its values of momentum and
density, divided the density range into density bins, and integrated the
momentum over each density bin, obtaining the distribution showed  in Fig.\
\ref{fig:mom_distr_01}b. This figure shows that the typical densities of the
entrained material are much larger in the radiative case than in the
adiabatic case; furthermore, the total mass of the entrained material turns
out to be of the same order in both cases. We can understand this if we
note that the initial shocks driven by the adiabatic jets heat the
external gas, which consequently expands and thus lowers its density.
In the radiative case, cooling suppresses the expansion, leading in contrast
to a compression of the shocked material; the momentum acquired by the 
external medium resides largely in material whose density is up to ten
times higher than the initial external density. In addition we must
remember that this shock-induced compression can be very effective in the
radiative case since strong shock waves survive for long times, from
$t=4$ to $t=10$ (see Fig. \ref{fig:lapl}).

\subsection{Light jets ($\nu=10$)}

\begin{figure*}
{\includegraphics[width=\hsize,bb=20 400 600 690]{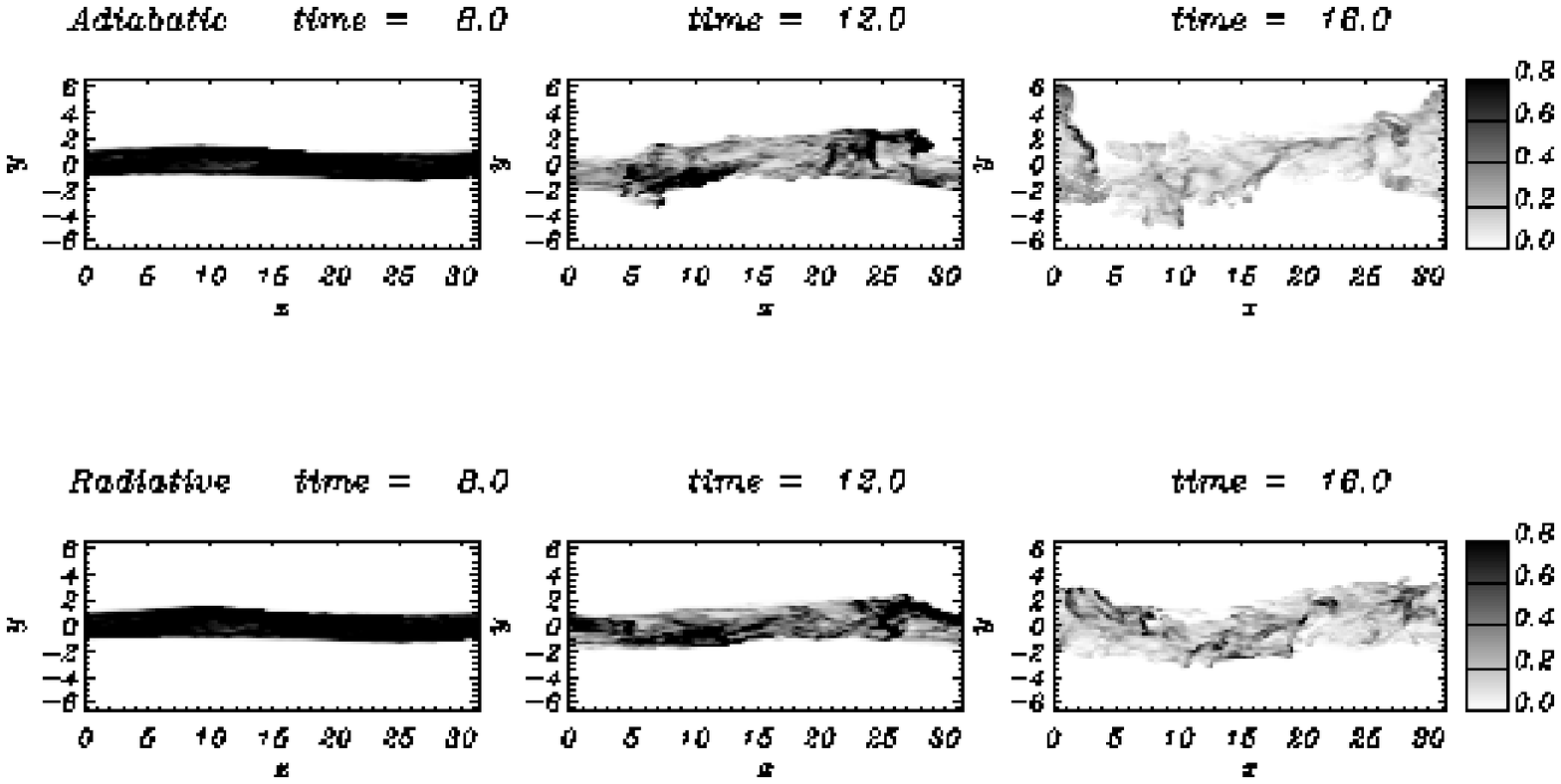}}
\caption{Grey-scale images of the jet density (i.e., of the product 
between density and tracer) distribution for light ($\nu=10$)
adiabatic (upper line) and radiative (lower line) jets; 
2D cuts at $z=0$.}
\label{fig:imdscl_10}
\end{figure*}

A quick glimpse at Fig.\ \ref{fig:image_10}, which displays \2d grey-scale
cuts of the density at different times in the $z=0$ and $x=5\pi$ planes,
shows that radiative losses in underdense jets have little effect on the
instability evolution, which is slower with respect to the previously
examined cases: the linear phase of the instability (stage 1) lasts until
time $t=8$ when the first shocks appear in both the adiabatic and in the
cooling jets. The density distributions shown in Fig.\ \ref{fig:image_10}
cannot show details inside the jet because the grey-scale saturates there.
Therefore, in order to obtain better views of the jet structures, we show
in Fig.\ \ref{fig:imdscl_10} grey-scale \2d cuts of the product of the
tracer and the density at different times. From these figures we can see that
shocks form inside the jet and in the external medium, corresponding to the
kinks induced by the fundamental helical mode, and persist for long times:
at time $t=16$, when mixing has taken place and the jet diameter has
considerably widened, a few weak shocks can still be detected both in the
adiabatic and cooling jets.

The growth of the higher order fluting modes is delayed in time, and
mixing occurs later, mainly because of the inertia of the external medium.
The onset of mixing occurs between times $t=10$ and $t=12$, and is
initially limited to the more external layers of the jet. At later
times (e.g., time $t=18$ in Fig.\ \ref{fig:image_10}) the action of 
mixing is not disruptive, and the jet maintains its collimation.

\begin{figure}[htbp]
{\includegraphics[width=\hsize,height=9.6cm,bb=90 90 570 720]{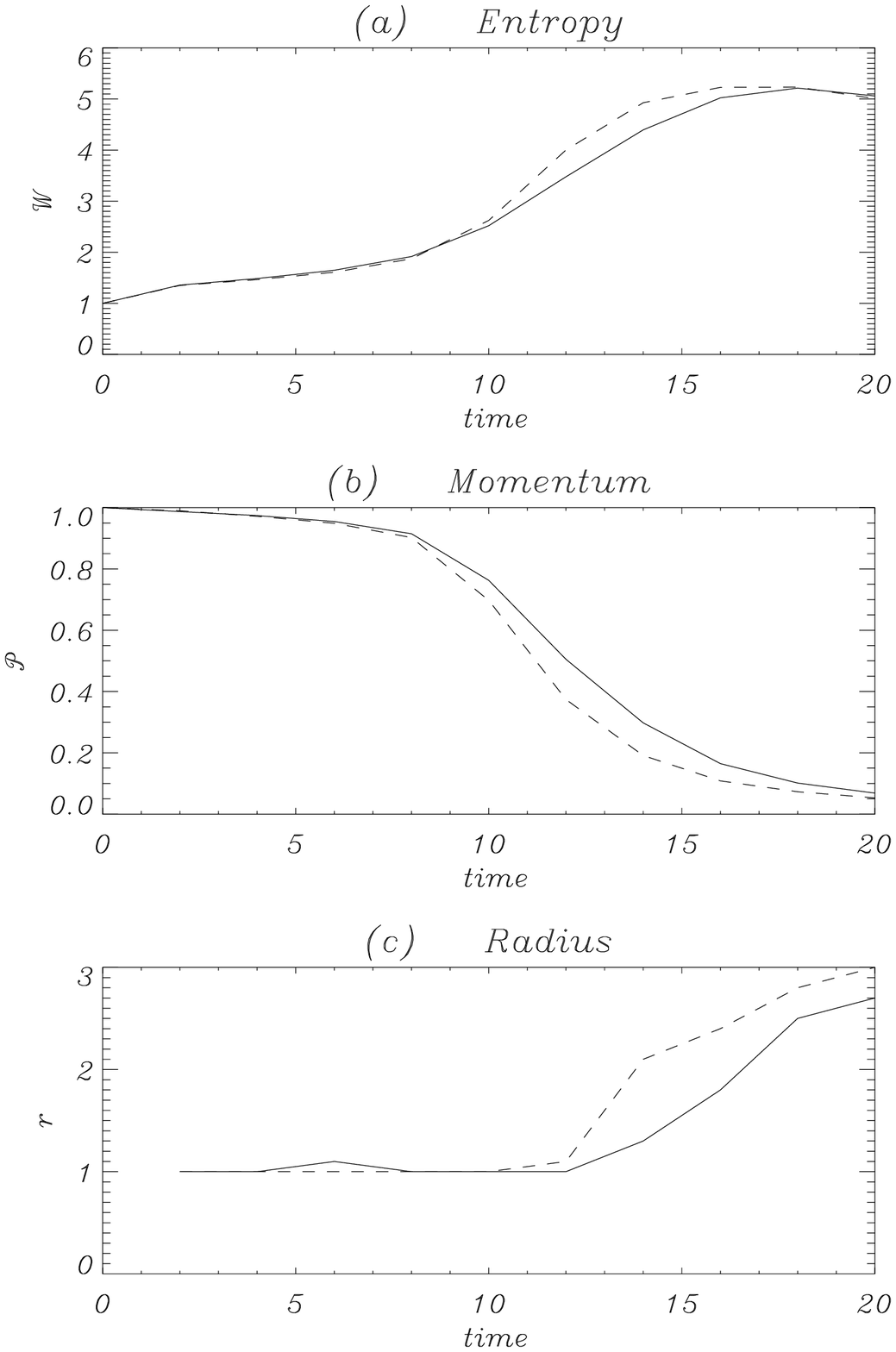}}
\caption{In panel (a) we show a plot of the tracer entropy for the radiative 
(solid line) and adiabatic (dashed line) jets as a function of time; 
in panel (b) we show a plot of the jet momentum for the radiative 
(solid line) and adiabatic (dashed line) jets as a function of time.
In panel (c) the temporal evolution of the jet radius for the
radiative (solid line) and adiabatic (dashed line) cases is plotted.
All the plots are for the light jet case ($\nu=10$).}
\label{fig:emr10}
\ef

The transition between the evolutionary stages is well depicted by the tracer
entropy, plotted in Fig.\ \ref{fig:emr10}a): entropy remains almost constant,
increasing slowly for the first 8-10 sound crossing times, when the
perturbations grow linearly and the first shocks form. A larger increase
of the entropy occurs at later times, with a steeper growth for the
adiabatic case (which reaches the quasi-stationary phase earlier, at $t
\sim 15$ as compared to $t \sim 18$ for the radiative case). The jet
momentum is plotted in Fig.\ \ref{fig:emr10}b) as a function of time; we
can see that the momentum behavior is similar to the entropy behavior:
the drop in the jet momentum appears to be steeper for the adiabatic
case, but it stops earlier so that the total amount of momentum lost
by the jet material is about the same for the two cases ($\sim$ 90\% of 
the initial value). A more careful comparison between the behaviors of
entropy and momentum shows that the phase of steeper decrease of momentum
begins earlier than the phase of steeper increase of entropy. In this case
we then have a distinct phase 2 (acoustic phase) of the evolution in which
momentum is mainly lost through shocks.   

\begin{figure}[htbp]
 
{\includegraphics[width=\hsize,bb=90 280 570 720]{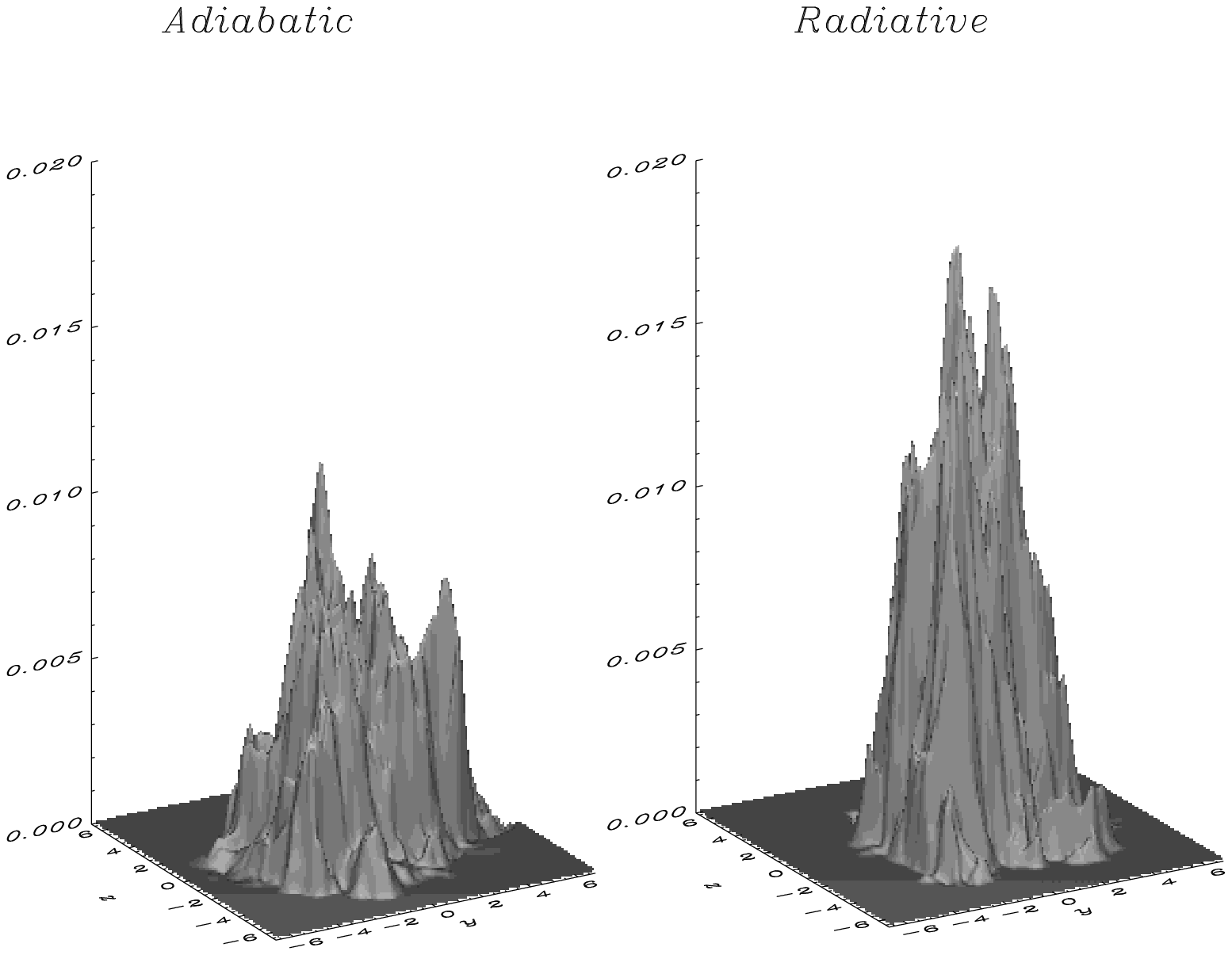}}

\caption{Longitudinally averaged distribution of the jet momentum
at time t=18 for the adiabatic and radiative light jets.}
\label{fig:avmom}
\end{figure}

The largest differences between the adiabatic and the radiative cases
can be seen in the last stage of the evolution: looking at Figs.\
\ref{fig:image_10} and \ref{fig:imdscl_10} at times $t=16$ and $t=18$, we
see that the adiabatic jet appears to be wider than the radiative one.
This is confirmed by Fig.\ \ref{fig:emr10}c), which shows the behavior of the
jet radius, as defined in the previous subsection, as a function of time.
The difference between the adiabatic and the radiative cases however is not
so large as for the dense jets.  This same effect can also be seen in the
jet momentum distribution: in Fig.\ \ref{fig:avmom} we show the grey-scale
distribution in the (yz) plane of the average of the jet momentum over the
longitudinal direction for the two cases; we see that the distribution
is wider for the adiabatic jet. Since the total amount of momentum retained
by the jet is about the same in the two cases, we also find that the maximum
value of the jet momentum, located at a position corresponding to the
initial jet axis, is smaller for the adiabatic jet (about half the maximum
value of the cooling jet momentum). 

From the above description we can conclude that -- in contrast to the heavy
jet case -- the presence of radiative losses does not strongly modify the
overall evolution of the instability in a light jet. To investigate the
reasons for this behaviour, we plotted in Fig.\ \ref{fig:loss} the total
radiative power (jet plus ambient medium) for heavy, light and equal-density
jets. In each case, the peak in the emission takes place shortly after the
onset of the acoustic stage, when shocks are well developed and not yet
disrupted by mixing. 

\begin{figure}[htbp]

{\includegraphics[width=\hsize,bb=100 460 550 700]{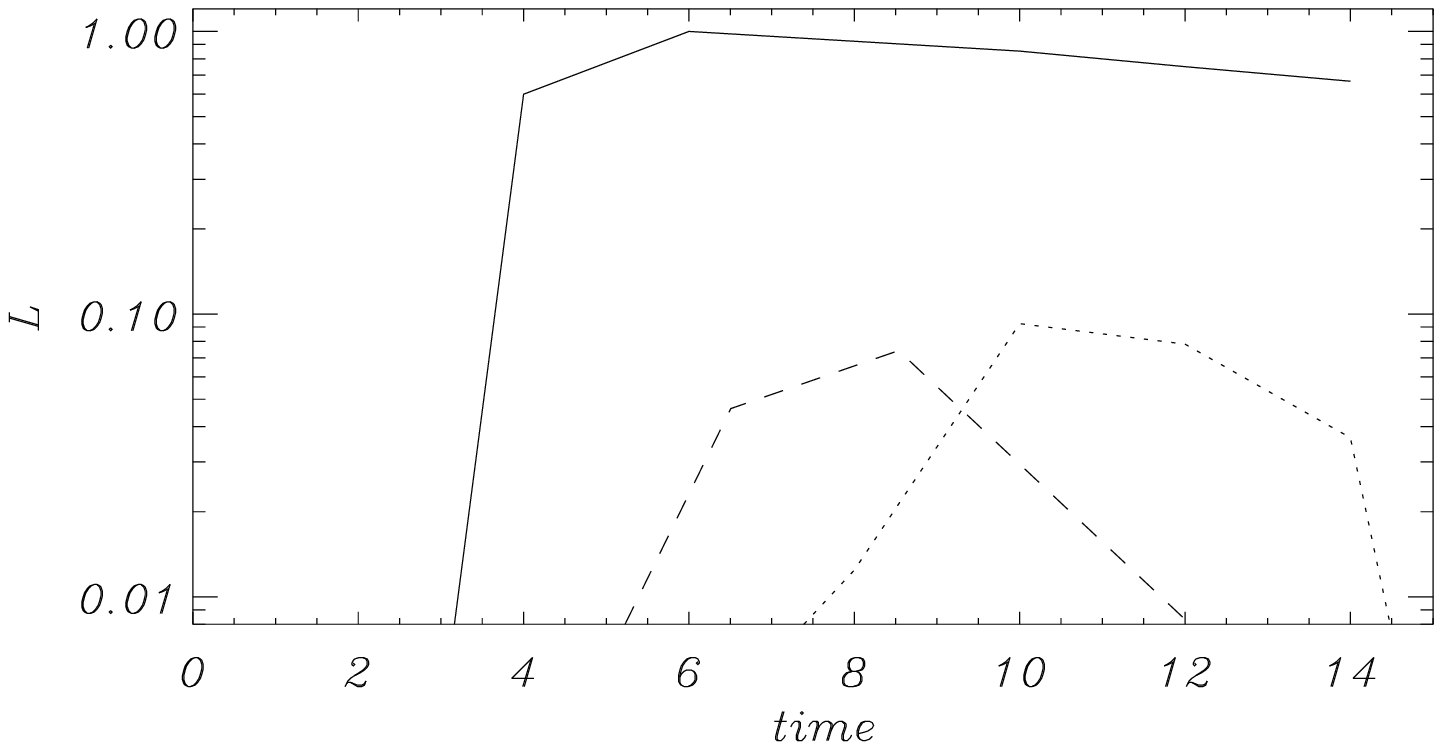}}
 
\caption{Power lost through radiation 
by heavy (solid line), light (dotted line) and equal-density 
jets (dashed line). 
Units are normalized to the 
peak emission in the $\nu=0.1$ jet.}
\label{fig:loss}
\end{figure}

The amount of energy lost through radiation is at least an order of
magnitude smaller in the light jet compared to the dense jet. The reasons 
for this behavior are many; first, the growth of the helical mode in the
dense jet drives larger amplitude kinks when compared to those observed in
the light jet; in this way the shock fronts are wider in the $\nu=0.1$
jet, and a larger amount of material is compressed, heated and ultimately
cooled through radiation. In addition, the wavelength of the most unstable
mode may play an important role: a shorter wavelength, typical of dense
jets (see Hardee \& Norman 1988) leads to the formation of a higher number
of shocks in a short time, increasing the efficiency of the cooling 
mechanism at shocks (compare also Figs.\ \ref{fig:image_01} and
\ref{fig:image_10}). We also note that the ambient medium gives a
significant contribution to the subtraction of energy from the system, in
particular when dense and equal-density jets are concerned; in fact, shocks
are driven by the jet into the ambient medium, and they propagate in the 
external regions immediately surrounding the jets; the spatial 
extension of these shocks can be consistently larger that in the jet's body.

\subsection{Equal-density jets ($\nu=1$)}

When studying equal-density radiative jets with the same value of $n_0 a$ as
considered in the previous cases ($n_0 a = 5 \times 10^{17}$ cm$^{-2}$,
consistent, for example, with a jet radius $a=5 \times 10^{15}$ cm and a jet
particle density $n_0=100$  cm$^{-3}$), we find that radiation introduces 
little effect on the instability evolution, as was the case for light jets.
This can be seen by comparing the first two columns of Fig.\ 
\ref{fig:long_cut_1}, where we show two-dimensional cuts of density in 
the plane $z=0$ for the adiabatic and radiative jets.

\begin{figure*}

{\includegraphics[width=\hsize,bb=20 250 600 730]{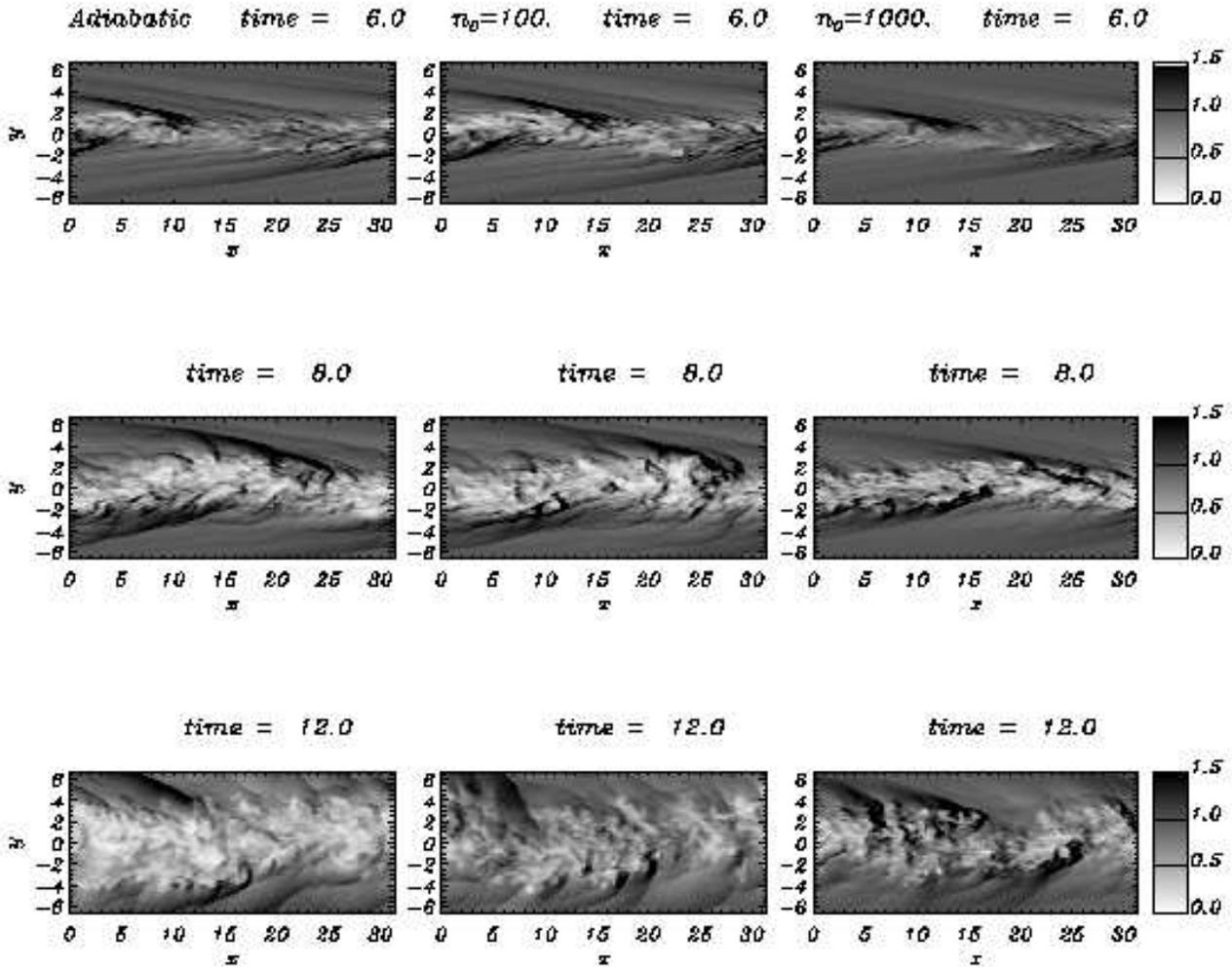}}

\caption{Grey-scale images of the density distribution for equidense
($\nu=1$) adiabatic (left column), and radiative jets, with the product of
the jet density and radius $n_0 a = 5 \times 10^{17}$ cm$^{-2}$ (center
column) and $n_0 a = 5 \times 10^{18}$ cm$^{-2}$ (right column);  2D cuts at
a fixed  $z$ ($z=0$).}
\label{fig:long_cut_1}

\end{figure*}

The smallness of the effects introduced by radiation could have been
predicted from the analysis of Fig.\ \ref{fig:loss}, which shows that the
amount of energy lost by radiation is similar to the light jet case, and is
smaller by more than an order of magnitude when compared to losses
associated with the heavy jet. For this reason, we have examined an
additional case with $n_0 a$ increased by an order of magnitude, i.e., $n_0 a
= 5 \times 10^{18}$ cm$^{-2}$, which is consistent, for example, with a jet
radius $a=5 \times 10^{15}$ cm and a jet particle density $n_0=1000$ 
cm$^{-3}$. This will increase radiative losses; for example, the amount of
energy lost by radiation at time $t=8$ corresponds to 0.08\% of the kinetic
energy of the jet at that same time, compared to a value of  0.02\% obtained 
in the previous case. The results for this high-cooling case
are also shown in Fig.\ \ref{fig:long_cut_1}, in the third column.

Looking at the general evolution, the upper panels of Fig.\
\ref{fig:long_cut_1} show that at time $t=6$ shocks have formed, as a
consequence of the non-linear growth of the helical mode; the fluting
modes drive deformations of the jet surface and entrainment of the
external medium  by the jet. (Previous times are not represented since
$t=6$ is the first time when non-linear effects can be detected.)\ \ The
evolution of fluting modes, which are responsible for early mixing, is
similar in equal-density jets and in heavy jets, for both adiabatic and
radiative cases: however, mixing between the jet and the external medium 
starts later (at time $t=6$) with respect to the $\nu=0.1$ case.
Mixing has an immediate and disruptive effect: external material is
entrained and the jet diameter widens quickly; at time $t=12$ the jet
maintains its large-scale structure and collimation, but on small 
scales it appears completely mixed with the ambient medium. Comparing the
results for the three cases, we see that the radiative cases show stronger
density enhancements inside the jets, and that the widening in the radiative
cases is reduced (this reduction is particularly evident in the
high-cooling case). This is again confirmed by Fig.\ \ref{fig:emr1}c), which
shows the behavior of the jet radius, as defined above, as a function of
time.

As for the heavy jet case, the acoustic and mixing stages cannot be
distinguished in the jet evolution, both for the adiabatic and radiative
cases. The onset of mixing is contemporary with the formation of the first
shocks in the flow, but the interaction of the two phenomena is different if
cooling is significant: in the adiabatic and $n_0=100$ jets, vigorous 
mixing takes place and shocks are destroyed after a few sound-crossing times, 
while in the
cooling $n_0=1000$ jet strong shocks survive up to the latest stages in the
jet evolution, when the global collimation of the jet has been partially
destroyed. In fact, in the adiabatic jet a strong shock appears only at
time $t=6$, while no shocks survive after time $t=8$ when a weak
discontinuity in the flow can be traced. 
In the cooling $n_0=1000$ jet,  instead, shock waves are not  destroyed
up to time $t=14$; this is approximately the time at which the cooling jet
enters the quasi-stationary stage of the evolution, as it is indicated by
the behavior of the  tracer entropy shown in Fig. \ref{fig:emr1}a). Entropy
grows more steeply for the adiabatic case: the quasi-stationary phase is
attained earlier, i.e., at time $\sim t=12$.

\begin{figure}[htbp]
{\includegraphics[width=\hsize,height=9.6cm,bb=90 90 570 720]{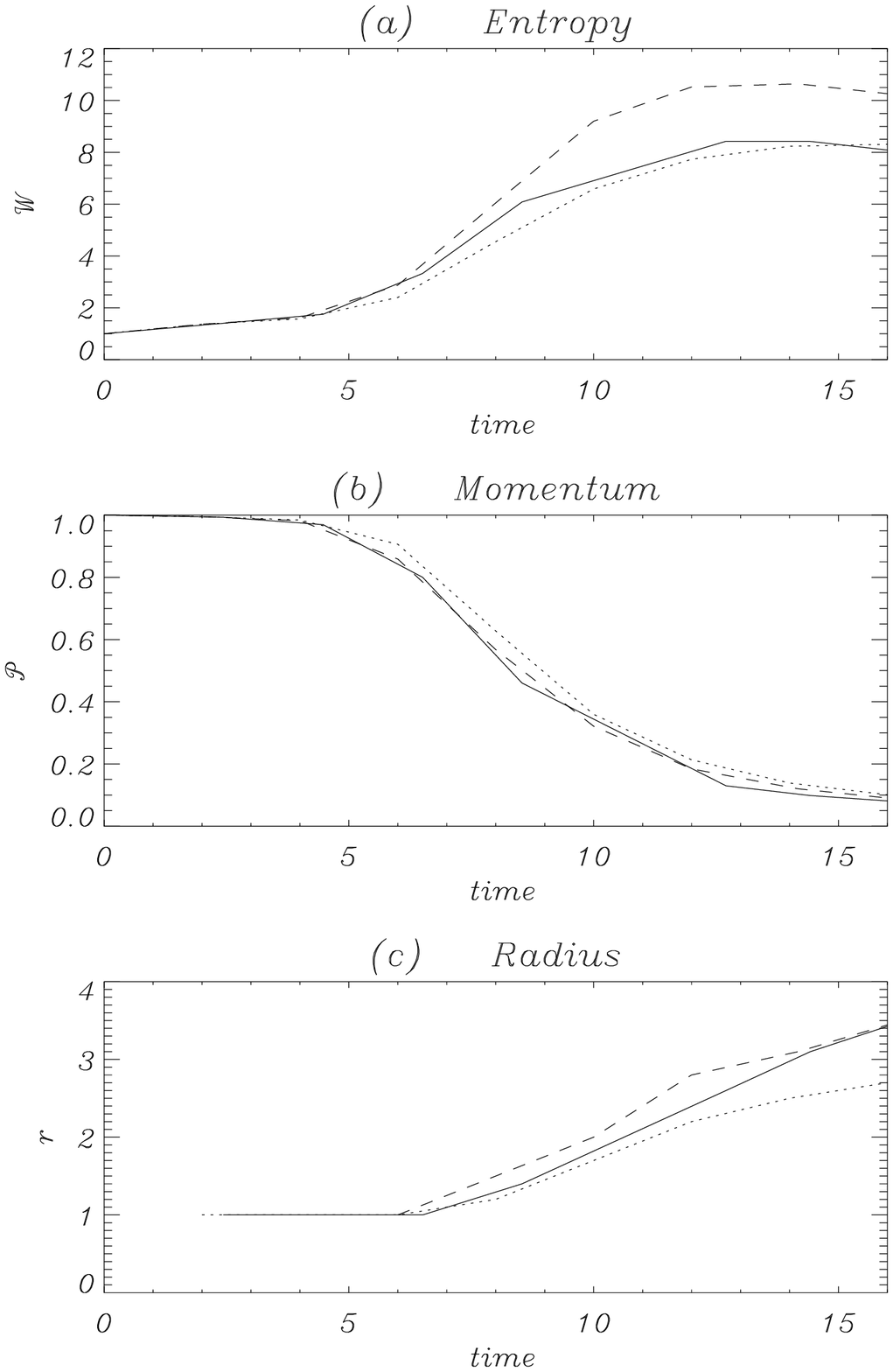}}
\caption{In panel (a) we show a plot of the tracer entropy for the radiative, 
$n_0 = 100$ (solid line), radiative, $n_0 = 1000$ (dotted line) 
and adiabatic (dashed line) jets as a function of time; 
in panel (b) we show a plot of the jet momentum for the radiative, 
$n_0 = 100$ (solid line), radiative, $n_0 = 1000$ (dotted line) and
adiabatic (dashed line) jets as a function of time. In panel (c) the
temporal evolution of the jet radius for the radiative,  $n_0 = 100$ (solid
line) , radiative, $n_0 = 1000$ (dotted line) and adiabatic (dashed line)
cases is plotted. All the plots are
for the equal density jet case ($\nu=1$).}
\label{fig:emr1}
\ef

Fig.\ \ref{fig:emr1}b) shows the variation in jet momentum as a function
of time for the radiative and adiabatic jets. The momentum loss appears to
occur slightly faster for the adiabatic and $n_0=100$ jets, although 
the final loss attained at the end of the evolution is similar for the
three cases, and corresponds to the $\sim$ 10\% of the initial jet momentum.
Although the integrated values are similar in all cases, the spatial
distribution of the jet momentum reflects the differences in the jet widening
noted above. 

\section{Energy spectra}

\begin{table*}[htb]

\caption{Value of the maximum kinetic energy in the transverse motions,
time at which this maximum is reached and  time at which equipartition
between longitudinal and transverse energy is attained, for all cases, 
and in four different scale intervals. Blank spaces are left where
equipartition was not observed within the maximum evolutionary time reached in
our calculations.}  

{\centering \begin{tabular}{|c|c|c|c|c|c|c|c|c|c|c|c|c|}
\hline 
&
\multicolumn{3}{|c|}{ $ \lambda <0.5 $}&
\multicolumn{3}{|c|}{$ 0.5 <  \lambda <1 $}&
\multicolumn{3}{|c|}{$ 1< \lambda <2 $ }&
\multicolumn{3}{|c|}{$ \lambda >2 $ }\\
\cline{2-4} \cline{5-7} \cline{8-10} \cline{11-13} 
&
$E_{max} $ &
$t_{max} $ &
$t_{eq} $ &
$E_{max} $ &
$t_{max} $ &
$t_{eq} $ &
$E_{max} $ &
$t_{max} $ &
$t_{eq} $ &
$E_{max} $ &
$t_{max} $ &
$t_{eq} $\\
\hline 
\hline 
$  \nu =0.1 $  - Ad. & 
$ 5.5 \times 10^{-3}$      & $6.$ & $9.5$    &
$ 6 \times 10^{-3} $       & $6.$ &          &
$ 6 \times 10^{-3} $       & $6.$ &          & 
$ 1.8 \times 10^{-2} $     & $6.$ &          \\
\hline 
$  \nu =0.1 $  - Rad. &
$  5 \times 10^{-3}  $     &   $6.$   &        &
$  5.5 \times 10^{-3} $    &   $6.$   &        &
$  4.5 \times 10^{-3} $    &   $6.$   &        &
$  8   \times 10^{-3} $    &   $6.$   &        \\
\hline 
$  \nu =1. $  - Ad. &
$ 10^{-2}            $     &   $8.1$  &  $11.5$  &
$ 10^{-2}            $     &   $8.5$  &  $13.$   &
$ 10^{-2}            $     &   $8.5$  &  $14.5$  &
$ 2.3 \times 10^{-2} $     &   $10.$  &          \\
\hline 
$  \nu =1. $  - Rad.&
$ 8. \times 10^{-3}  $     &   $8.1$  &  $12.5$  &
$ 8. \times 10^{-3}  $     &   $8.8$  &  $14.$   &
$ 7. \times 10^{-3}  $     &   $8.8$  &  $16.$   &
$ 1.6 \times 10^{-2} $     &   $10.$  &         \\
\hline 
$  \nu =10. $  - Ad.&
$ 1.3 \times 10^{-2} $     &   $12.$  &  $13.2$  &
$ 1.3 \times 10^{-2} $     &   $12.$  &  $16.$   &
$ 10^{-2}            $     &   $13.$  &  $18.$   &
$ 2. \times 10^{-2}  $     &   $14.$  &          \\
\hline 
$  \nu =10. $  Rad.&
$ 10^{-2}            $     &   $12.$  &   $16.$  &
$ 8 \times 10^{-3}   $     &   $12.$  &   $18.5$ &
$ 8 \times 10^{-3}   $     &   $13.$  &   $20.$  &
$ 1.5 \times 10^{-2} $     &   $14.$  &          \\
\hline 
\end{tabular}\par}

\label{tabak}

\end{table*}

We have seen that radiation losses tend to maintain jet collimation. To
obtain a different perspective on the instability evolution and on the
differences between the various cases, we have analyzed the relative
amounts of kinetic energy in longitudinal and transverse motions 
at different scales.
 
When the instability grows, the kinetic energy of the longitudinal jet bulk
motion goes into internal energy, radiation, and kinetic energy in the
transverse motions. In the direction perpendicular to the jet axis we can
have: a) large-scale motions, due to the growth of the helical
unstable modes, and b) small-scale motions, induced by the \KH fluting
modes, and by a possible turbulent cascade from large to small scale
motions. Our aim is to analyze in more detail the transfer of energy from
longitudinal to transverse motions; the key issue is to identify possible
differences due to radiative losses. This analysis can also give us
informations about the persistence of the jet as a coherent structure of
longitudinal velocity.

We perform this analysis by calculating the three-di\-men\-sio\-nal Fourier spectra
of the kinetic energy for longitudinal and transverse motions. We then
obtain one-di\-men\-sio\-nal spectra by integrating first over the transversal 
wa\-ve\-num\-bers $k_y$ and $k_z$, and second over the longitudinal and azimuthal
wave numbers (after transforming to cylindrical coordinates). We use these
one-dimensional Fourier spectra ${\rm A}_x(k_x)$ and ${\rm A}_r(k_r)$ to
compare the amount of energy in the longitudinal and radial direction, at
all times, and for all the studied cases. Initially, before the jet enters the
non-linear phase, the energy is found in
longitudinal motions at large scales; as the instability evolves, a portion
of the kinetic energy is transferred to longitudinal motions at smaller
scales and to transverse motions. In Fig.\ \ref{fig:escale} we can follow
this process for the particular case of underdense jets. The figure shows the
behavior of longitudinal and transverse energies, normalized to the 
initial jet kinetic energy, as a function of time for
different scale ranges (recall that lengths are given in units of the jet
radius). Looking at the longitudinal energy at large scales ($\lambda > 2$,
panel a) we see that this component, as noted above, initially represents the
only form of energy present in the system and, as the instability evolves,
this component decreases monotonically. At smaller scales (panels
b,c,d) the longitudinal energy first increases, reaches a maximum, and then
decreases. The time at which the maximum is reached is about the same for
all the scale ranges. A similar behavior is found for the transverse energy,
but the maximum is reached at later times. The faster decrease observed
in longitudinal energy when compared to that observed for the transverse
energy leads at some time to an equipartition between the two and, from then
on, motions can be considered to be isotropic. This isotropy is first
reached at small scales, and then at larger scales. We further note, by
comparing the curves for the adiabatic case and those for the radiative
case, that in this last case the transfer of energy to transverse motions
is less efficient and more energy is always kept in longitudinal motions. 

The behavior described for the underdense case is found also for the other 
cases; we have then summarized all our results in Table \ref{tabak}, where we
report, for all our cases and for each scale interval, the time at which
the maximum of transverse energy is reached, the value of this maximum, and
the time at which equipartition between longitudinal and transverse energy
is attained.

By comparing the table with the behavior of entropy and momentum shown above,
we can see that the time at which the maximum transverse energy is reached
falls in the steepest portion of the momentum drop and of the energy growth,
confirming the connection between the mixing process and the transfer of
energy to the transverse motions. In the case of radiative jets, this 
process of energy transfer is less efficient as demonstrated by the lower
values of $E_{max}$ for these cases and, as discussed above, mixing and
jet widening is lower for these types of jets. Comparing the different
values of the density ratio, we see that $E_{max}$ increases slightly as
$\nu$ is increased, i.e., going from overdense to underdense jets. This
increase however cannot compensate for the increase in density with $\nu$;
therefore we expect lower transverse velocities for the high $\nu$ cases, and
therefore a smaller jet widening, as it is indeed observed in our
simulations.
    
\begin{figure}

{\includegraphics[width=\hsize,bb=100 410 570 780]{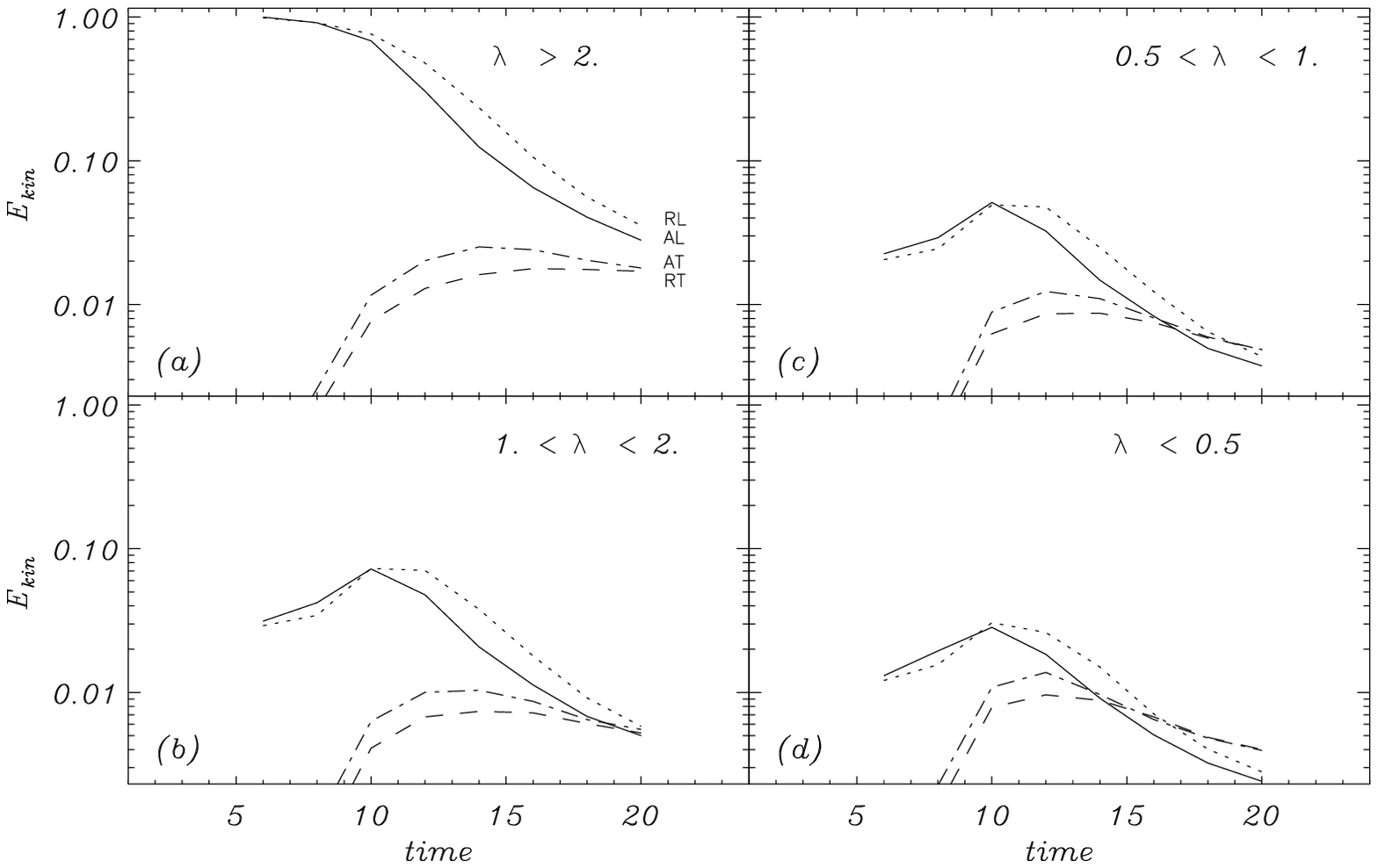}}

\caption{Kinetic energy for longitudinal and transverse motions, as a
function of time for light jets.
Solid and dotted lines represent the energy of longitudinal
motions respectively for adiabatic and radiative jets; dot-dashed and dashed
lines  represent the energy of transversal motions again for adiabatic and
radiative jets.}
\label{fig:escale}

\end{figure}

The table also shows that near the end of our simulations for the $\nu =1$
and $\nu = 10$ cases, we reach equipartition between the longitudinal and
transverse energies also at scales larger than the jet radius, while in the 
overdense case equipartition is found only at small scales. We can therefore
ask whether, in these cases, the jet can still be identified as a coherent
velocity structure. To answer this question, we have computed the
autocorrelation function for the longitudinal velocity, defined as 
$$
F(h) = \int v_x(x,y,z) v_x(x+h,y,z) dx dy dz
$$
where the integral is performed over the whole domain of integration.
We have then defined a correlation length as
$$
l_c = \int_0^L F(h)/F(0) dh
$$
and have plotted in Fig. \ref{fig:corrl} $l_c$ as a function of time for
all the different cases. The three panels in the figure refer respectively 
to the overdense case (upper panel), the equal density case (middle panel)
and the underdense case (lower panel); the solid curves are for adiabatic
cases and the dashed curves are for radiative cases.
  
\begin{figure}

{\includegraphics[width=\hsize]{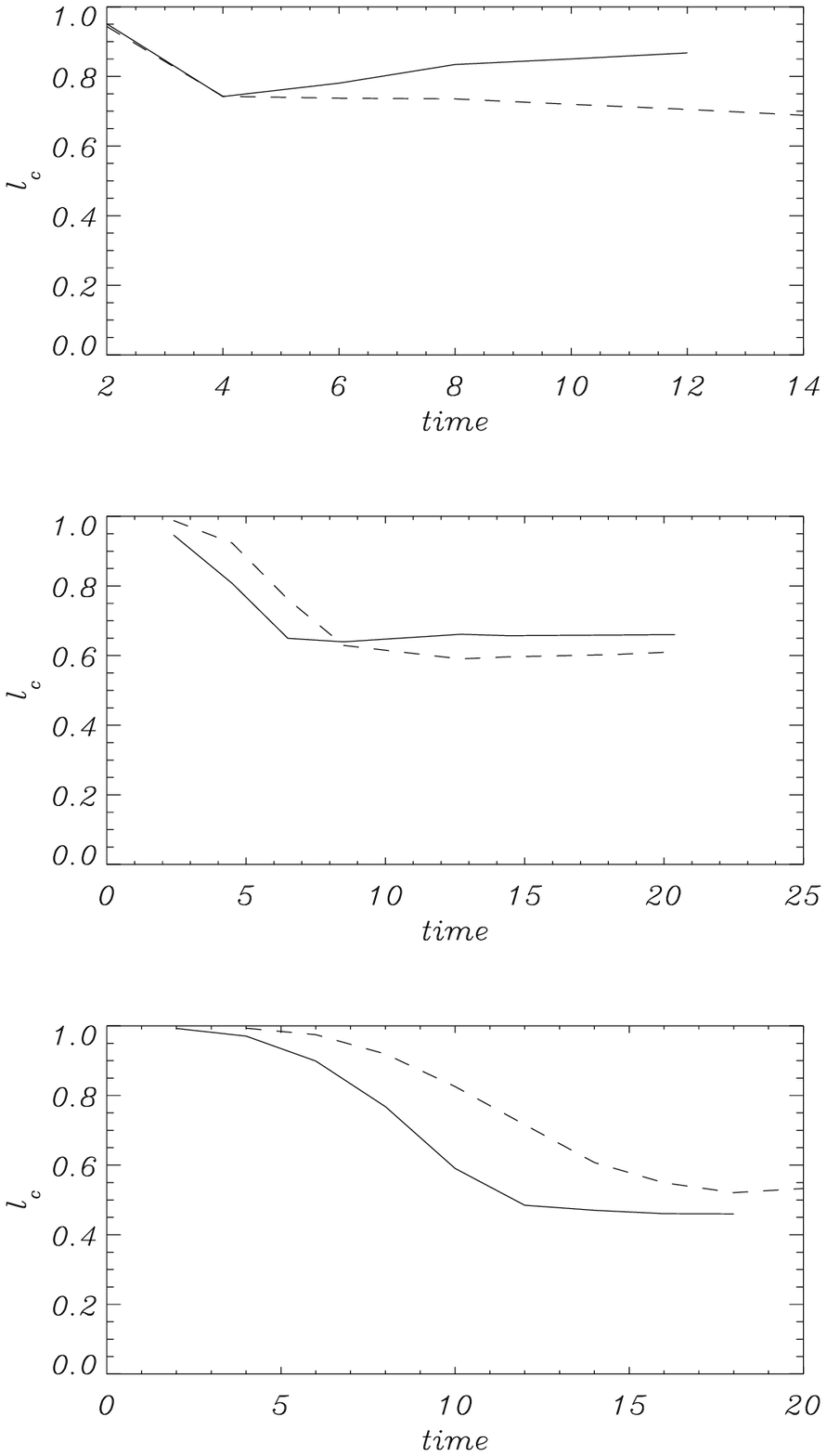}}

\caption{Longitudinal velocity autocorrelation length as a function of time,
the autocorrelation length is in units of the size of the domain. The upper
panel refers to $\nu=0.1$, the middle panel refers to $\nu = 1$ and the lower 
panel refers to $\nu = 10$. The solid curves are for the adiabatic cases, 
while the dashed curves are for the radiative cases. }
\label{fig:corrl}

\end{figure}

From Figure \ref{fig:corrl}, we see that, as was also suggested by the 
Fourier analysis described above, the overdense case maintained an higher 
coherence than the other two cases, especially compared to the 
low density case in which $l_c$ drops to $\sim 0.4$ at the end of the 
simulation. Radiation acts in different directions for the different values 
of $\nu$: in the overdense case radiation tends to give a more inhomogeneous
structure both in density (as we have already seen) and in velocity and
this is reflected by the lower value of $l_c$, in the equal density case the 
effects of radiation are very small, while in the underdense case radiation
tends to keep a more coherent structure with an higher value of $l_c$.  

\section {Summary and Conclusions}

We studied the evolution of the \KH instability in three-dimensional
jets, solving the hydrodynamical equations through a finite-differences
numerical code, adopting a temporal approach and varying the 
external-to-ambient medium density; we thus considered dense, underdense 
and iso-dense jets. The main results of our calculations can be summarized
as follows:
\begin{itemize}

\item {\bf Heavy jet}

Dense jets undergo fast evolution. Both helical and fluting modes grow
non-linearly at the earliest stages of the evolution, giving rise to large
scale deformations in the structure of the jet and to entrainment of
ambient material. If cooling is present, the duration of the acoustic phase
of the evolution is increased, and shocks survive in the flow for longer
times. Furthermore, the strength of the fluting modes is reduced; as a
result, mixing with the ambient medium is less effective and the jet
retains its collimation for longer times. The momentum loss in the
adiabatic and radiative cases is similar, but it occurs in different ways: in
both cases, the momentum is transferred away from the jet axis. However,
the adiabatic jet loses momentum by expansion and entrainment of ambient
medium far from the jet axis, material which is heated and expands by shocks
that are driven into the ambient matter from the jet; in contrast, the
radiative jet transfers its momentum to ambient material which is located
near the jet surface, material which has been shocked and cooled, and thus
condensed relative to its initial state.

\item {\bf Light jet}

For underdense jets the evolution develops over 20 sound crossing times, 
and thus is slower than the previous case. Mixing takes place later with
respect to the onset of the acoustic stage, and its action is damped by the
inertia of the external medium. The final jet widening is caused by the
growth in amplitude of the fundamental helical mode, which is the main
source of shock formation and momentum loss. The amount of energy lost
through radiation at a fixed time is smaller than in the previous case, but
cooling has nonetheless a small effect on the instability evolution, again
reducing the  amplitude of the jet widening.

\item {\bf Equidense jet}

Iso-dense jets evolve similarly to dense jets, although the onset of 
the various stages is delayed in time. The amount of energy lost through 
radiation is smaller, and the short duration of the shock-dominated stage
suppresses the effects of cooling on the overall instability evolution.
To demonstrate the effects of radiative losses, we ran a simulation 
in which we increased the jet particle density (which is one of the radiative
control parameters). The main effects of cooling are a longer duration of
shocks in the flow, damping of the helical unstable mode (and thus of the
amplitude of the jet deformations in the transverse direction) and of the 
fluting modes, which implies that mixing is limited to the external layers of
the jet.

\end{itemize}

These results indicate that radiative losses indeed cause jet stabilization,
an effect which is more evident in the overdense case and less evident in the
other cases; they are thus consistent with the results of previous 2D 
studies by Rossi et al.\ (1997) and Micono et al.\ (1998), even though the 
effect is not large as in the cases studied by these authors. This difference
can be explained as a result of the assumptions of (i) cylindrical geometry
with axial symmetry and (ii) axi-symmetric initial perturbations for the 2D
cases. The evolution of the instability in two dimensions is then dominated
by strong shocks on the jet axis: the amount of energy lost through
radiation in these strong compressions is high, and the stabilizing effects
of cooling are enhanced. In our 3D study, the strength of the shock waves
that form is smaller, both because of geometry and because of small-scale
motions driven by the growth of the \KH fluting modes (the latter are
neglected in two dimensions). In this way, the effects of cooling are weaker,
and their importance depends on the density ratio between the jet and the
ambient medium. However, our results can confirm that cooling is efficient
in reducing the disruptive action of the instability, in maintaining the jet
collimation for longer times, and in extending the temporal duration of the
stage where shocks are present in the flow.

\bigskip

{\it Acknowledgments:} The calculations have been per\-for\-med on the Cray T3E 
at CINECA in Bologna, Italy, thanks to the support of CNAA. 
This work has been supported in part by  and by the DOE ASCI/Alliances grant at the
University of Chicago. M.M. acknowledges the CNAA 3/98 grant. 

\section{References}

\noindent
Birkinshaw, M. `Beams and Jets in Astrophysics' 1991

\noindent
Bodo, G., Massaglia, S., Ferrari, A., Trussoni, E., 1994, A\&A 283, 655 

\noindent 
Bodo, G., Massaglia, S., Rossi, P., Rosner, R., Malagoli, A., Ferrari, A., 
1995, A\&A 303, 281

\noindent 
Bodo, G., Rossi, P., Massaglia, S., Ferrari, A., Malagoli, A., 
Rosner, R., 1998, A\&A 333, 1117

\noindent
Colella, P., Woodward, P.R., 1984, J.\ Comp.\ Phys.\ 54, 174 

\noindent
Downes, T.P., Ray, T.P., 1998, A\&A 331, 1130

\noindent
Ferrari, A., 1998, Ann.Rev.As.Ap. 36, 539

\noindent
Gerwin, R.A., 1968, Rev.Mod.Phys. 40, 532

\noindent
Hardee, P.E., Norman, M.L., 1988, ApJ 334, 70

\noindent
Hardee, P.E., Clarke, D.A., Howell, D.A., 1995, ApJ 441, 644

\noindent
Hardee, P.E., Clarke, D.A., 1995, ApJ 451L 25

\noindent
Hardee, P.E., Clarke, D.A., Rosen, A., 1997, ApJ 485, 533

\noindent
Hardee, P.E., Stone, J.M., 1997, ApJ 483, 121

\noindent 
Massaglia, S., Trussoni E., Bodo G., Rossi P., Ferrari A. 1992,
A\&A 260, 243

\noindent
Micono, M., Massaglia, S.,  Bodo, G., Rossi, P., Ferrari, A., 1998,
{\it A\&A} {\bf 333}, 989

\noindent
Rossi P., Bodo, G., Massaglia, S., Ferrari, A., 1997, A\&A
321, 672  

\end{document}